\documentclass[aps,twocolumn,floatfix,byrevtex,titlepage,balancelastpage,raggedbottom,amsfonts,amssymb,amsmath,showkeys]{revtex4} 
\usepackage{epsfig}
\usepackage{bm}
\usepackage{times}
\usepackage{mathptmx}
\usepackage{amsmath}
\usepackage{graphicx}
\usepackage{graphics}
\usepackage{amssymb}
\usepackage{bm}
\usepackage[english]{babel}
\begin{document}
\title{Lattice Boltzmann simulation of the surface growth effects for the infiltration of molten Si in carbon preforms}
\author{Danilo Sergi}
\affiliation{University of Applied Sciences (SUPSI), 
The iCIMSI Research Institute, 
Galleria 2, CH-6928 Manno, Switzerland}
\author{Loris Grossi}
\affiliation{University of Applied Sciences (SUPSI), 
The iCIMSI Research Institute, 
Galleria 2, CH-6928 Manno, Switzerland}
\author{Tiziano Leidi}
\affiliation{University of Applied Sciences (SUPSI), 
The iCIMSI Research Institute, 
Galleria 2, CH-6928 Manno, Switzerland}
\author{Alberto Ortona}
\affiliation{University of Applied Sciences (SUPSI), 
The iCIMSI Research Institute, 
Galleria 2, CH-6928 Manno, Switzerland}

\date{\today}


\begin{abstract}
The infiltration of molten silicon into carbon preforms is a widespread technique employed
in the industry in order to enhance the thermal and mechanical properties of the final ceramic
products. A proper understanding of this phenomenon is quite challenging since it stems from the 
reciprocal action and reaction between fluid flow, the transition to wetting, mass transport, 
precipitation, surface growth as well as heat transfer. As a result, the exhaustive modeling 
of such problem is an involved task. Lattice Boltzmann simulations in 2D for capillary infiltration 
are carried out in the isothermal regime taking into account surface reaction and subsequent surface 
growth. Precisely, for a single capillary in the linear Washburn regime, special attention is paid 
to the retardation for the infiltration process induced by the thickening of the surface behind the 
contact line of the invading front. Interestingly, it turns out that the process of surface growth
leading to pore closure marginally depends on the infiltration velocity. We conclude that porous matrices
with straight and wide pathways represent the optimal case for impregnation. Our analysis includes
also a comparison between the radii characterizing the infiltration process (i.e., minimum, hydraulic,
average and effective radii).
\end{abstract}
\maketitle

\section{Introduction}

The Lattice Boltzmann (LB) method is gaining increasing consideration for 
hydrodynamic simulations. A variety of systems can be studied in the
incompressible limit \cite{book1,book2,wolfram,review}. The advantage 
of the LB method over other numerical schemes resides in its ability to deal with non-equilibrium
dynamics, interface phenomena (wetting) and complex geometries (porous media). Our aim
is to simulate the infiltration of molten silicon (Si) into carbon (C) preforms,
taking into account the reaction of silicon carbide (SiC) formation 
\cite{hillig,mortensen,dezellus,dezellus2,passerone,highT,nisi,nisi2,htc09,einset1,einset2,messner}. 
The benefits of this process mainly have to do with the thermal properties. 

In the simulations, the chemical reaction responsible for surface growth is 
based on mass precipitation \cite{euromat_narciso,reaction_pre,dissolution,crystal,d2q4,snow}.
The effects caused by the thickening of the surface are analyzed for a single
capillary exhibiting the linear Washburn law \cite{washburn,chibarro,succi2}.
This time dependence for the penetration of the invading front is typical of 
reactive Si infiltration \cite{nisi2,htc09}. It is important to stress here
that the LB models reside on assumptions that are not fulfilled in experiments.
For example, in the simulations the liquid and vapor phases have almost the same density \cite{succi2}.
But this inconsistency is not dramatic since the density difference at the interface
determines the surface tension (Laplace law) and in turn the contact angle.
So, once the experimental contact angle of $30^{\circ}$ \cite{nisi,nisi2} is reproduced the results
can be regarded as reliable in that respect. The discrepancies with experimental data
are mainly due to the fact that in the simulations the Reynolds number overestimates
the experimental value, particularly low (micron scale and weak velocities) \cite{einset1,nisi2}. 
This means that matching with experimental results could be restored in the limit of capillaries 
of infinite length. However, the shortcomings arising from this inaccuracy should be limited 
since the reactivity is expected to be affected by the infiltration velocity not significantly
\cite{dissolution,crystal,snow}. In any case, our investigation is carried out under
severe simplifying conditions. Last but not least, thermal effects are disregarded \cite{si_thermal}.

With our simulations we can ascertain that surface growth retards the process of 
capillary infiltration. We are also in a position to discuss the process of pore closure.
The main result of our work is that the phenomenon of pore closure is to a large extent
independent of the infiltration velocities. Guidelines for optimal melt infiltration are 
formulated accordingly.

\begin{figure*}[t]
\includegraphics[width=12cm]{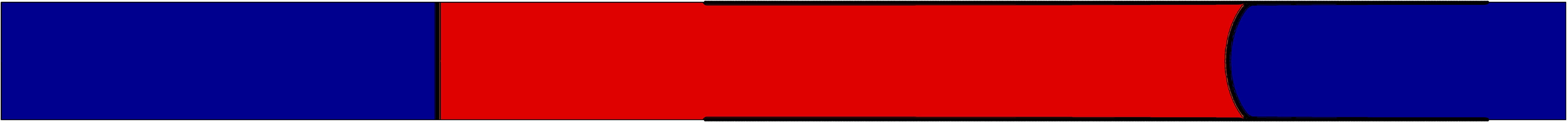}\\
\vspace{0.1cm}
\includegraphics[width=12cm]{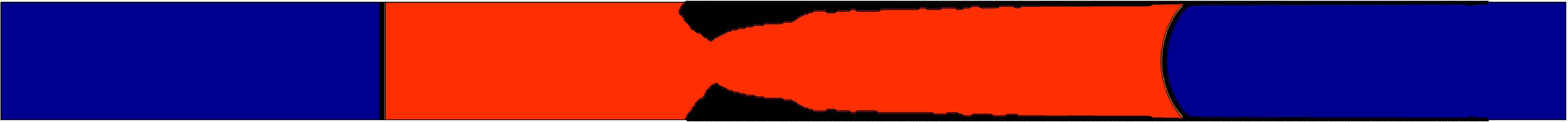}
\caption{\label{fig:capillary}
Fluid motion inside an interstice driven by capillary forces.
In red the main wetting component, in blue the other substance. The solid phase is represented by black
points. The contact angle is the angle that the interface (tangent) forms at the contact line, where the 
three phases meet. Top: case without surface reaction; bottom: outcome with reaction enabled.}
\end{figure*}
\begin{figure*}[t]
\includegraphics[width=8.5cm]{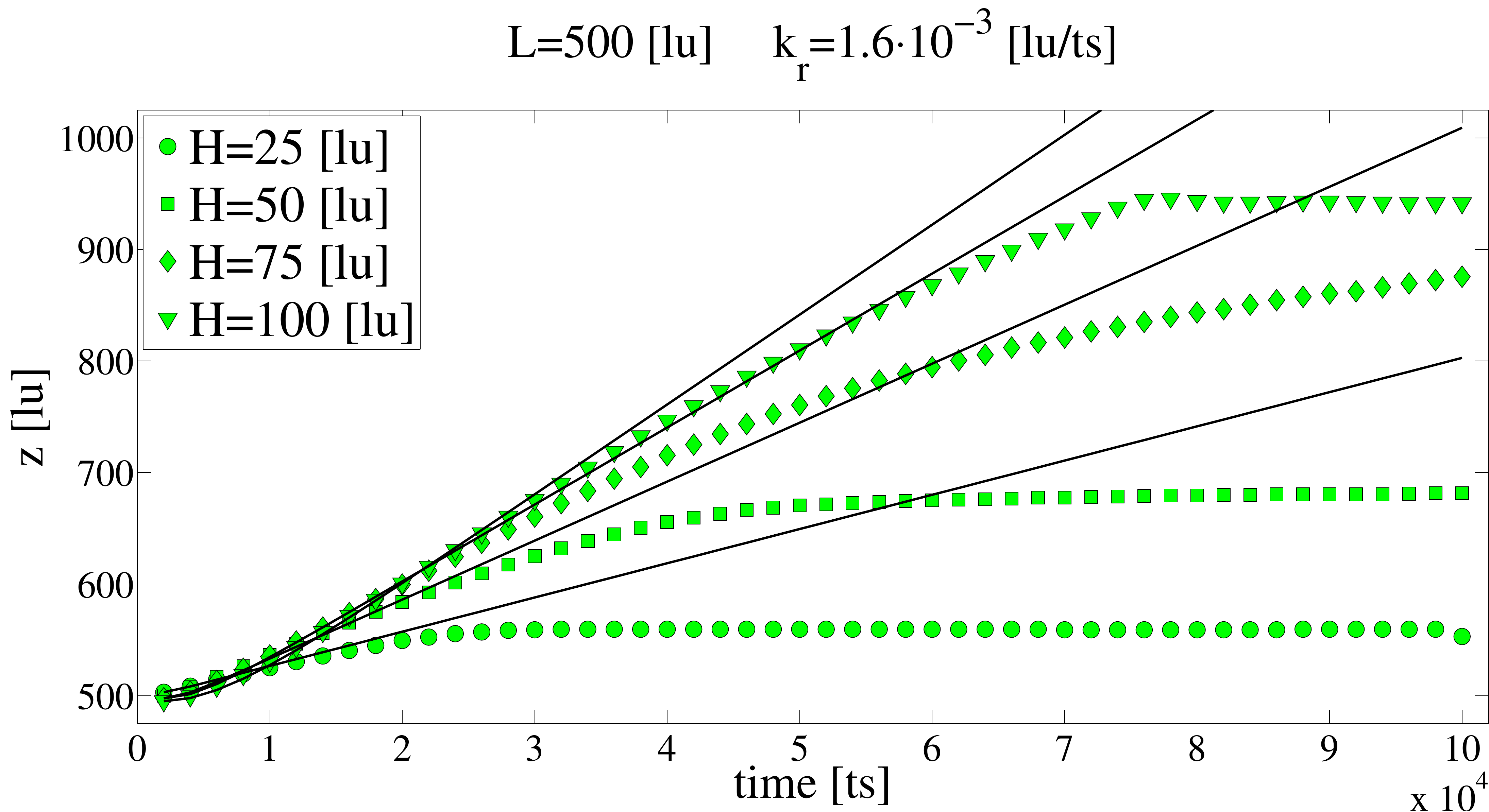}\hspace{0.5cm}
\includegraphics[width=8.5cm]{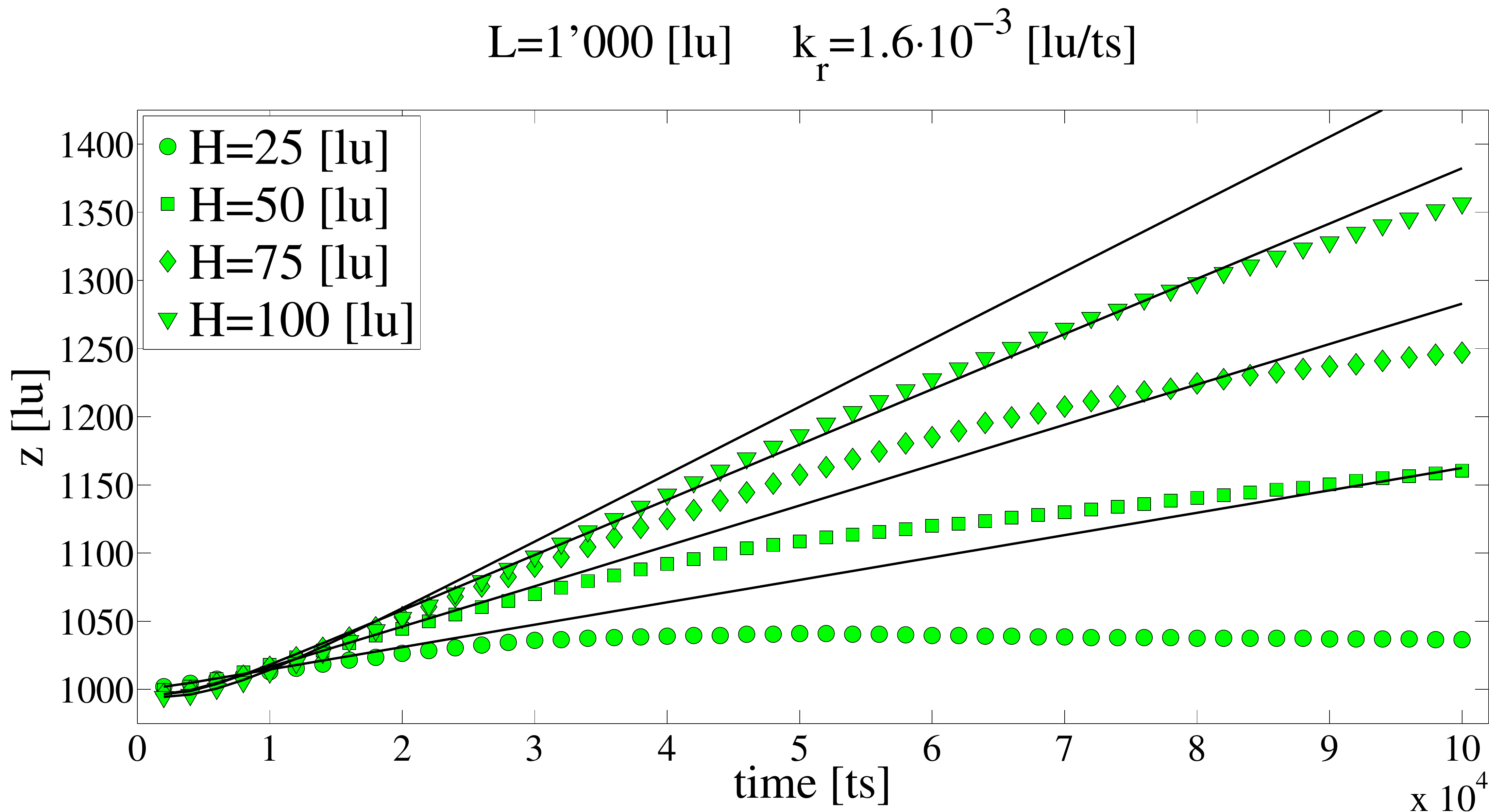}
\caption{\label{fig:front}
Centerline position of the invading front in the course of time. Green points represent the results obtained 
from simulations with reactive surfaces. The solid lines correspond to the theoretical predictions in the 
absence of reaction.}
\end{figure*}
\begin{figure*}[b]
\includegraphics[width=8.5cm]{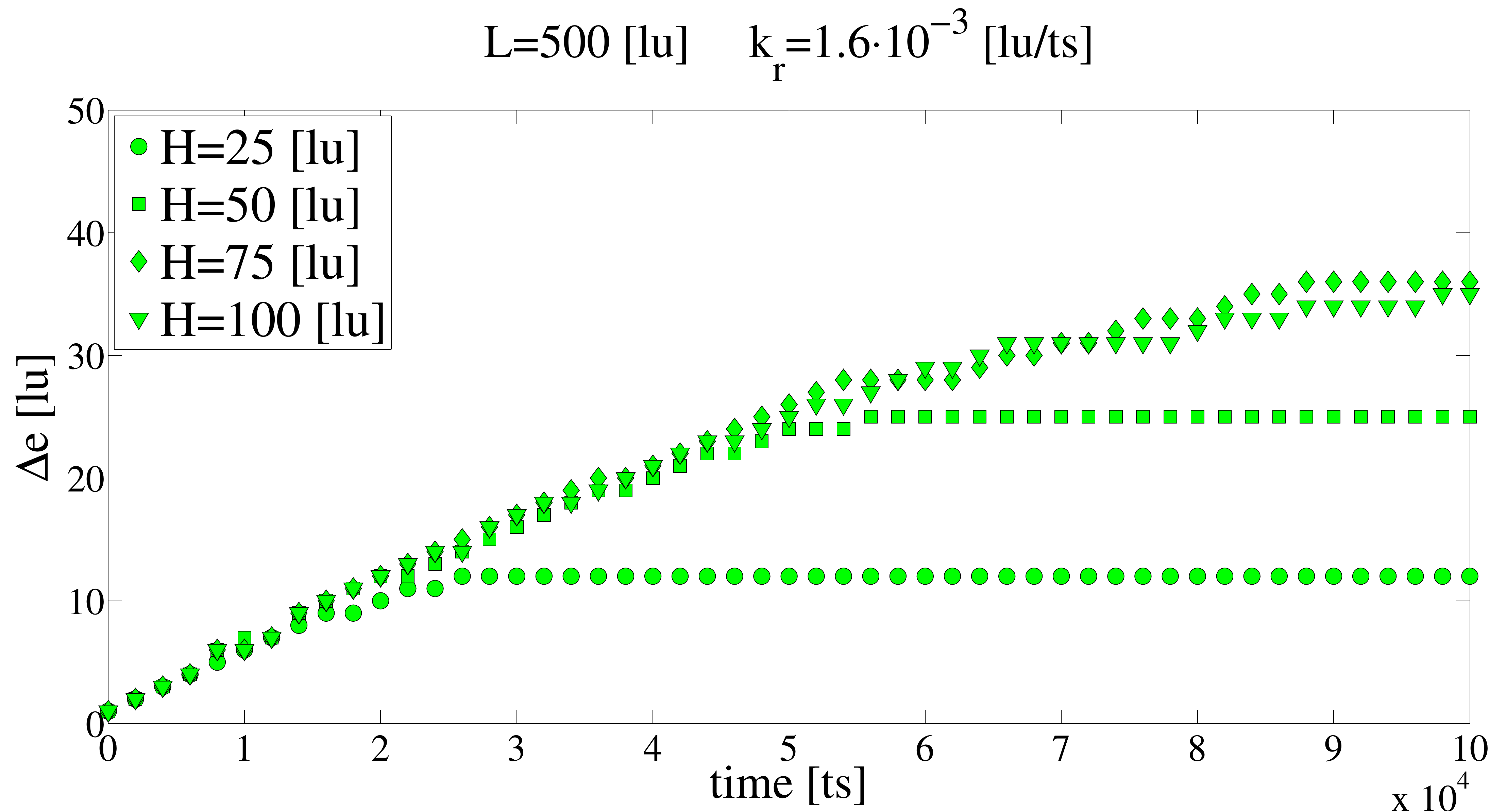}\hspace{0.5cm}
\includegraphics[width=8.5cm]{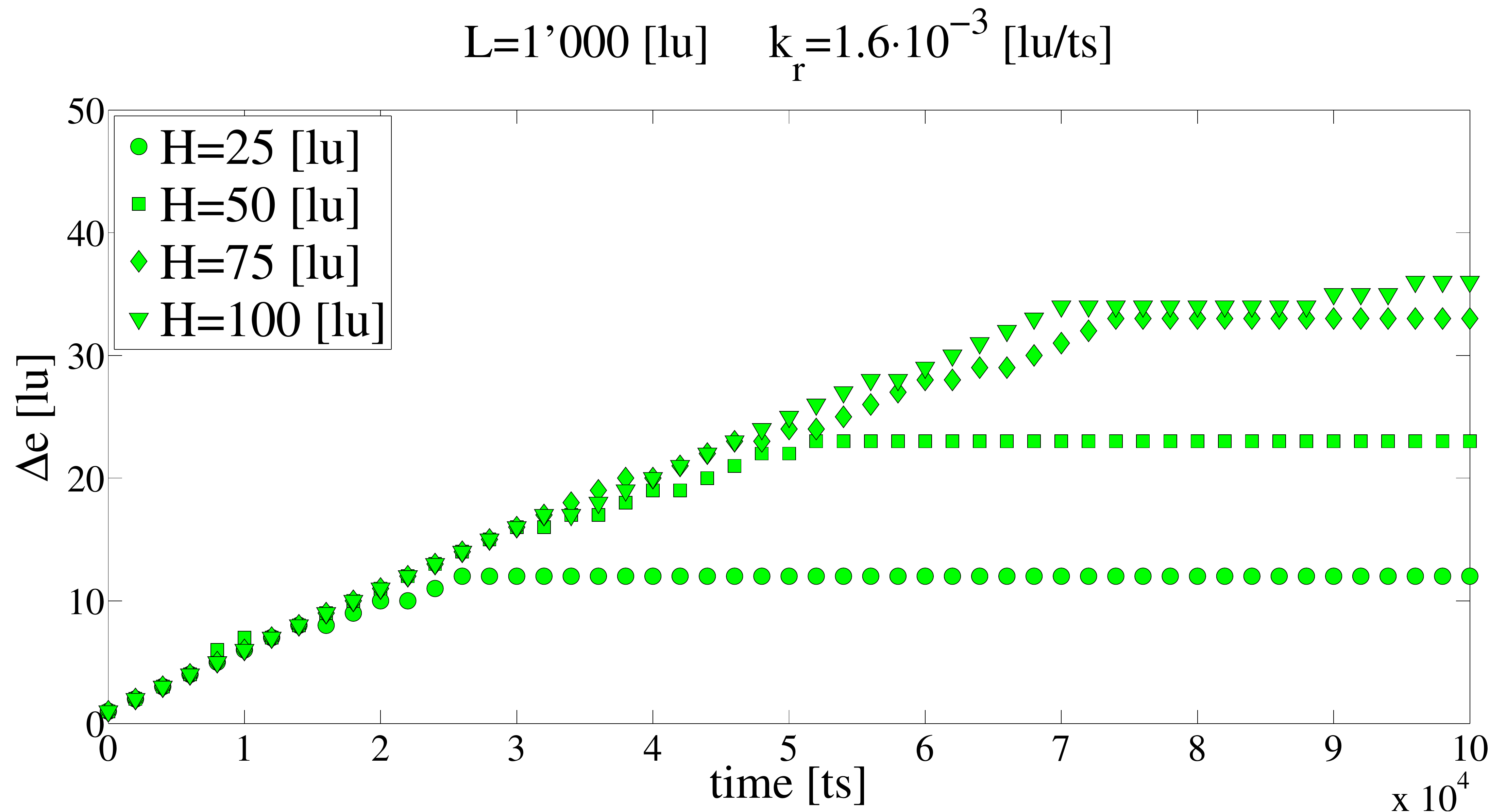}\\
\includegraphics[width=8.5cm]{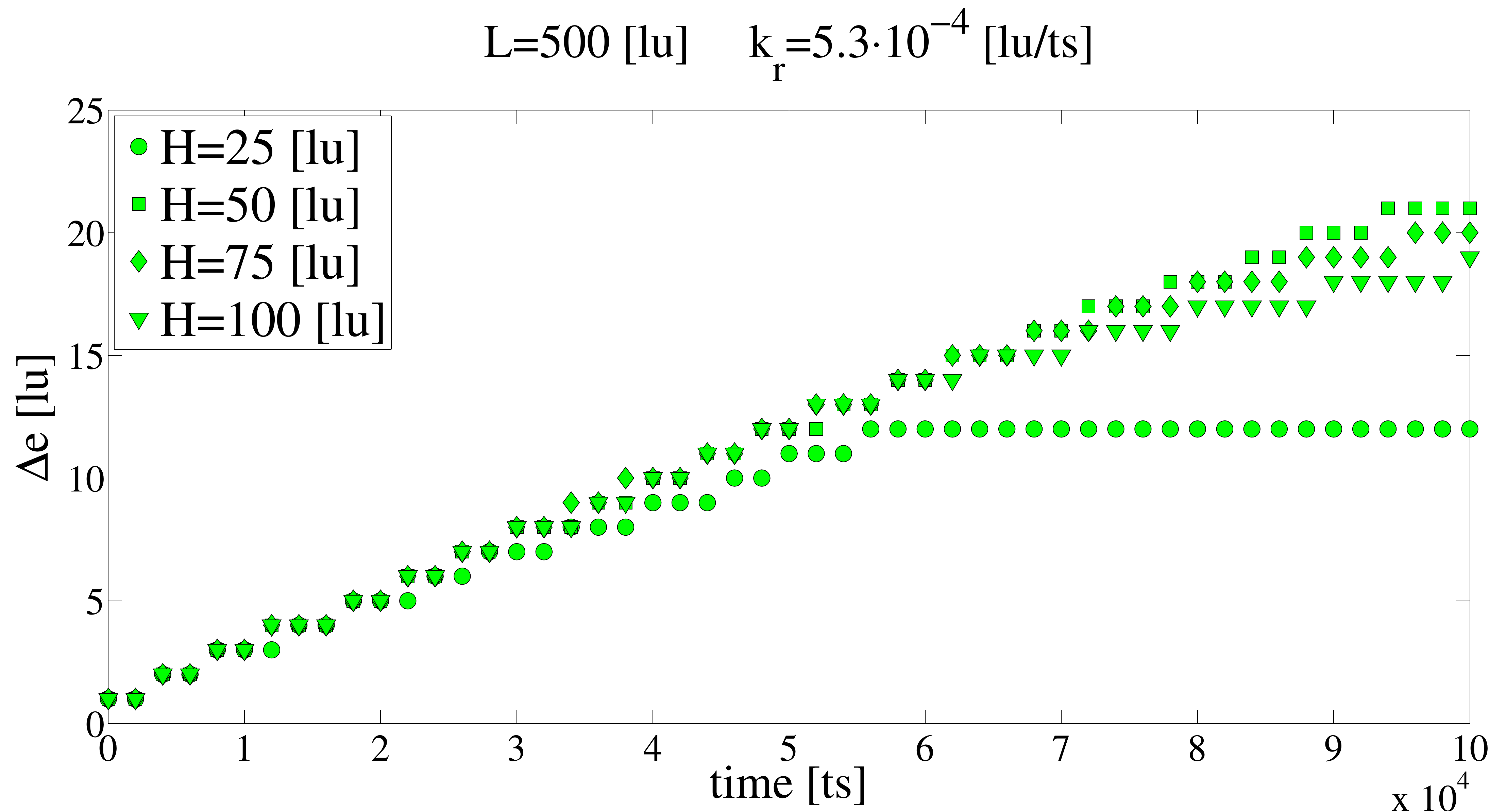}\hspace{0.5cm}
\includegraphics[width=8.5cm]{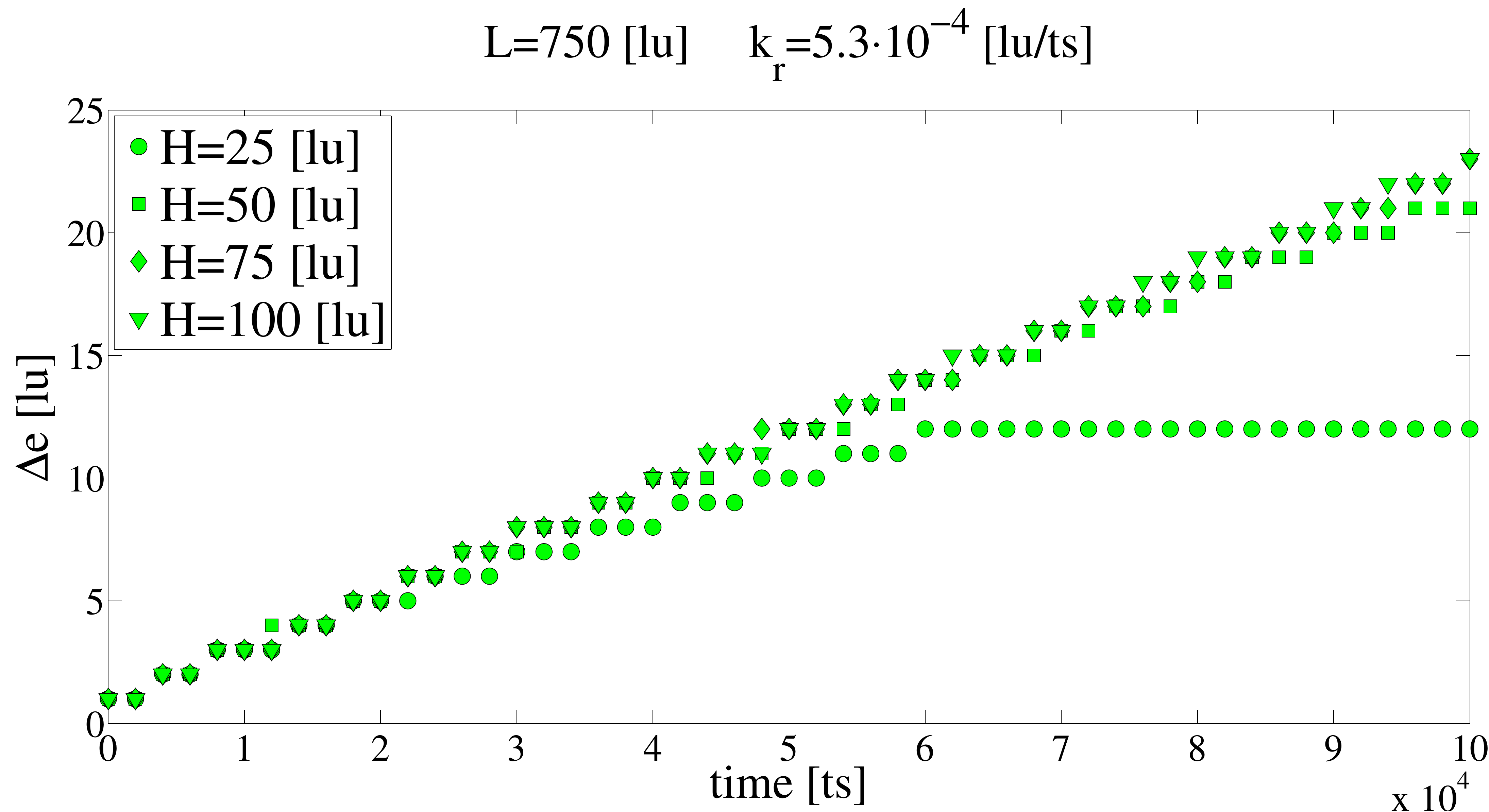}
\caption{\label{fig:solid}
Time dependence of the maximal width of the solid phase $\Delta e$.}
\end{figure*}
\begin{figure*}[t]
\includegraphics[width=8.5cm]{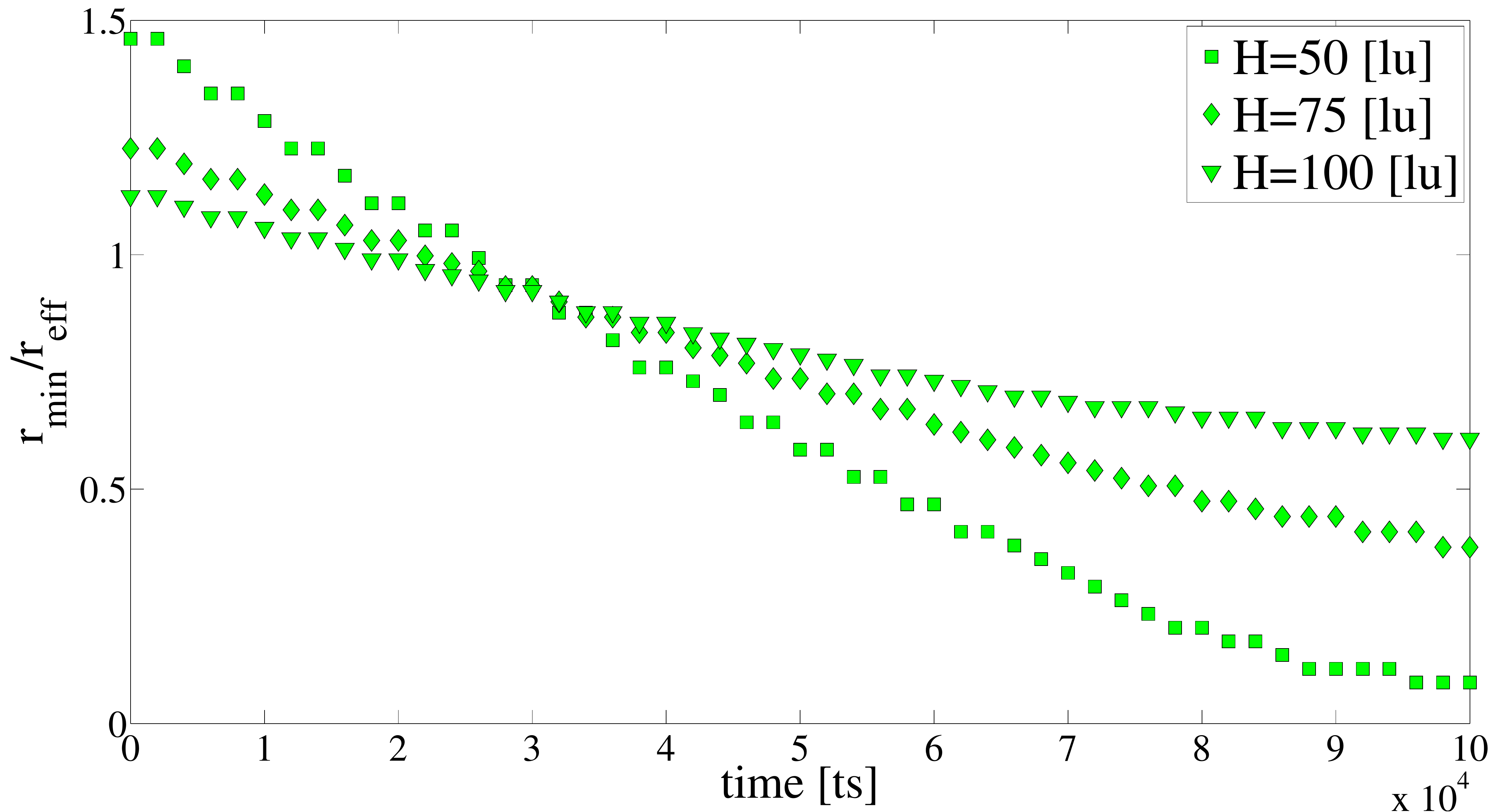}\hspace{0.5cm}
\includegraphics[width=8.5cm]{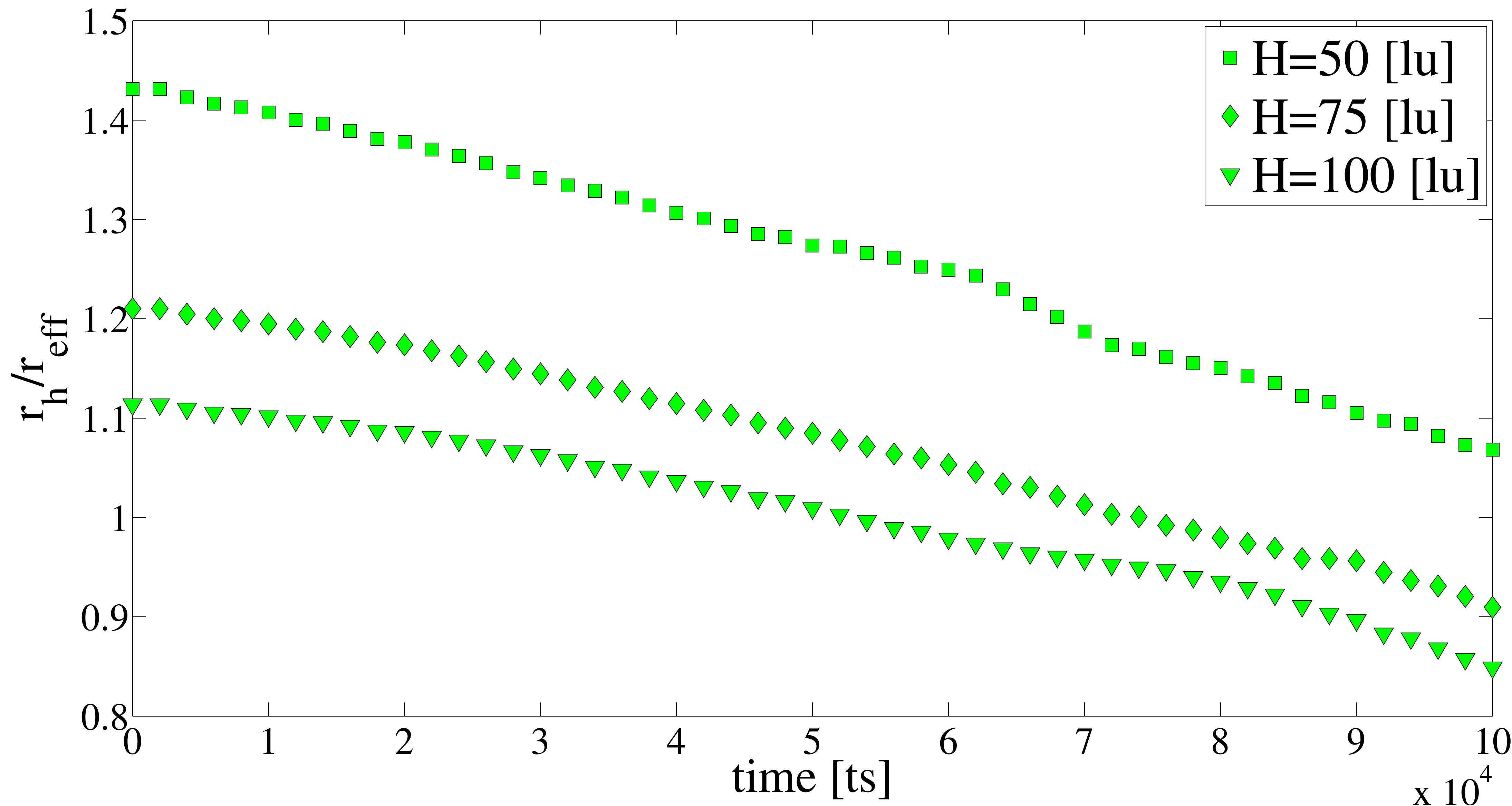}\\
\includegraphics[width=8.5cm]{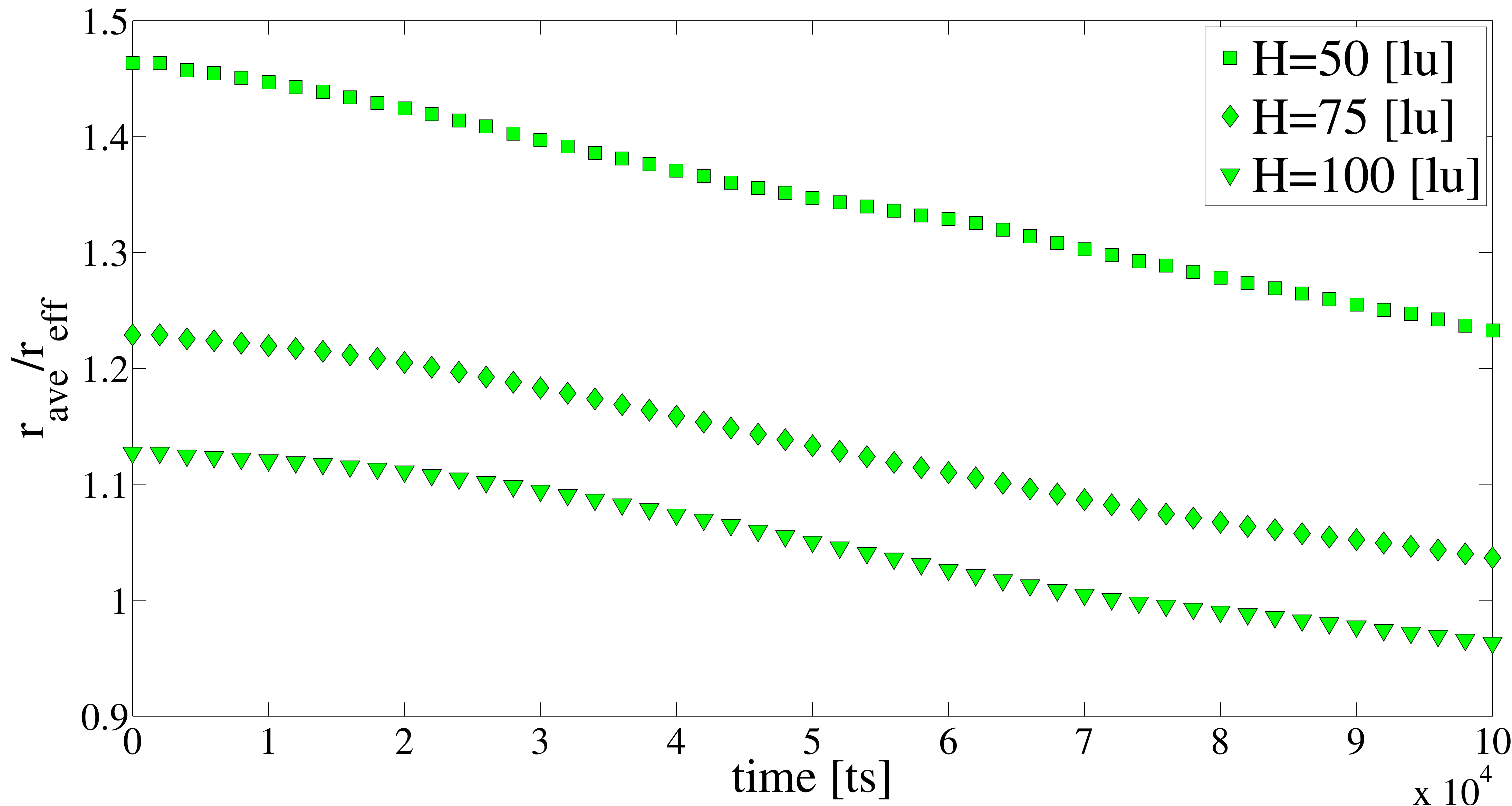}\hspace{0.5cm}
\includegraphics[width=8.5cm]{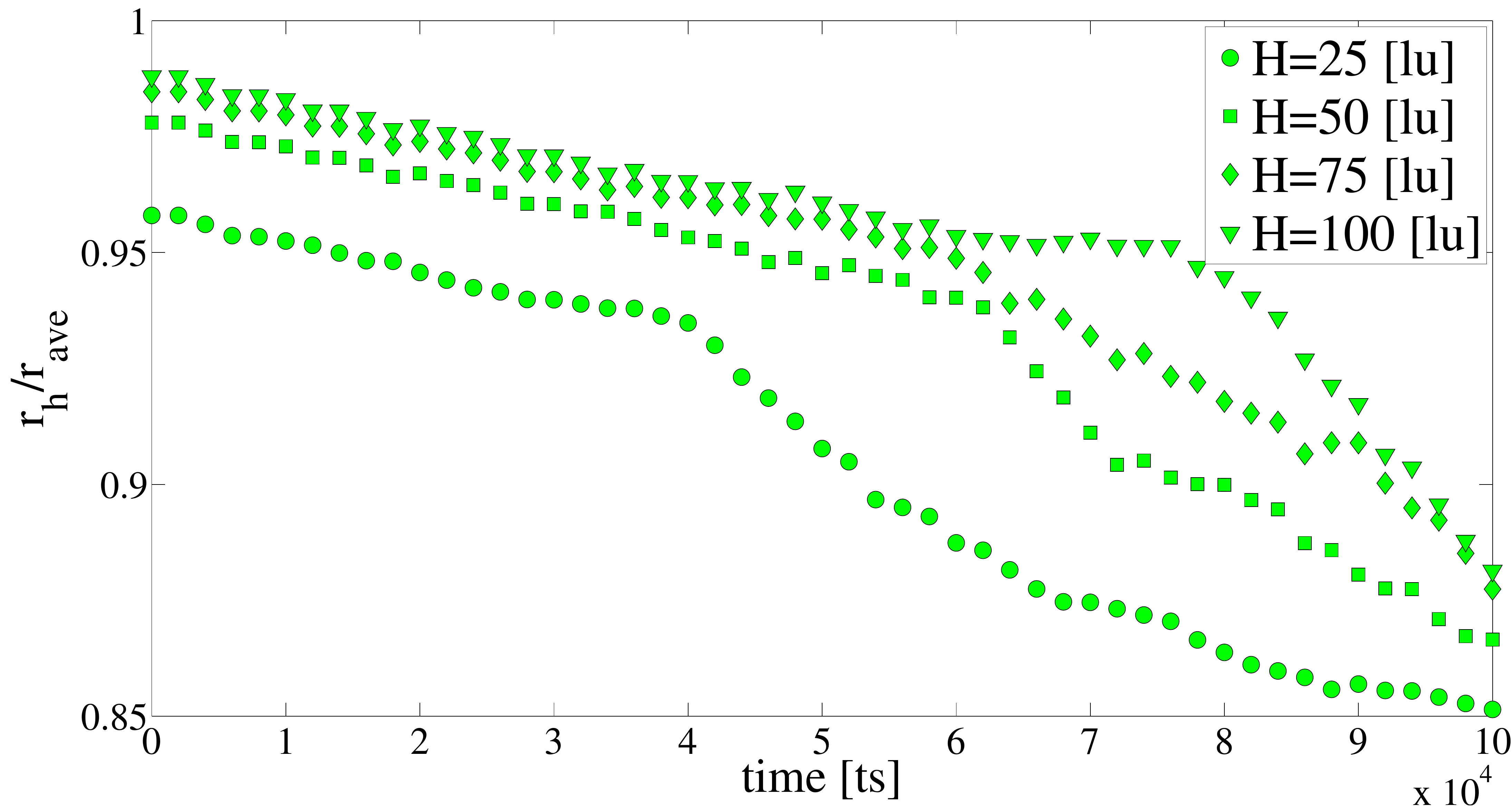}
\caption{\label{fig:radii}
Comparison between the minimum, hydraulic, average and effective radii, denoted respectively by $r_{\mathrm{min}}$, 
$r_{\mathrm{h}}$, $r_{\mathrm{ave}}$ and $r_{\mathrm{eff}}$. For all plots, $L=500$ [lu] and $k_{\mathrm{r}}=8\cdot 10^{-4}$ [lu/ts].
Similar curves are obtained for the other capillary lengths. For weaker reactivities, i.e.~smaller $k_{\mathrm{r}}$, the
various ratios are closer to $1$.}
\end{figure*}

\section{LB models}

The LB method is based on the discretization of the velocity space \cite{book1,book2,wolfram,review}, 
besides the discretization of the simulation domain, common to other numerical schemes. The LB method 
simulates the hydrodynamic behavior at a mesoscopic level since combining rules reminiscent of microscopic 
mechanisms and statistical treatment. The evolution of the systems is determined by iterating the BGK equation 
\cite{bgk}. The bounce-back rule is applied for the collisions between the fluid and solid phases. It has 
been proven that the physics of capillary infiltration is better described by multicomponent systems 
\cite{chibarro,succi1,succi2,succi3}. Interface phenomena (surface tension) are introduced by taking
into account fluid-fluid interactions (cohesive forces). Wetting phenomena can be accounted for by
incorporating solid-fluid interactions (adhesive forces). In our simulations, both types of interactions
are implemented using the models proposed in Ref.~\cite{martys}.

In the LB framework, mass transport is described under the assumption of weak coupling with fluid flow. 
Also in this case, the dynamics satisfies the BGK equation \cite{reaction_pre,dissolution,crystal,d2q4,snow}, 
in full analogy with fluid motion. The reaction at the surface is based on mass precipitation
\cite{reaction_pre,dissolution,crystal,d2q4,snow}: in the course of time mass deposits on the surface and when
the cumulated mass exceeds a threshold value, the solid phase expands. In our simulations, the relative effects
between reactivity and the characteristics of mass transport is varied only by considering different values of
the reaction-rate constant $k_{\mathrm{r}}$ as done in Ref.~\cite{capillary}. A thorough account on this 
aspect is proposed in Ref.~\cite{crystal}.

In the next section, we shall present the results for capillaries of different length, height and reaction conditions.
The simulations generate data in model units. The units of all quantities will be expressed in terms of the units 
for length lu, for mass mu and for time ts. Direct comparison with quantities given in ordinary units is possible
after suitable transformations \cite{snow,raabe_units}, even though it is better to establish equivalences via
dimensionless parameters like the Reynolds number \cite{landau}. The details about the LB models, the simulation settings
and related issues can be found in Ref.~\cite{capillary}.


\section{Results and discussion}

Our starting point is the linear Washburn law for capillary infiltration \cite{succi2}: 
\begin{equation}
z(t)=\frac{V_{\mathrm{cap}}H\cos\theta}{6L}t_{\mathrm{d}}[\exp(-t/t_{\mathrm{d}})+t/t_{\mathrm{d}}-1]\ ,
\label{eq:z}
\end{equation}
where $V_{\mathrm{cap}}=\gamma/\mu$ and $t_{\mathrm{d}}=\rho H^{2}/12\mu$. Furthermore, $z$ is the centerline 
position of the invading front, $\gamma$ the surface tension, $\mu$ the dynamic viscosity, $\rho$ the density, 
$H$ the capillary height, $L$ the length, $\theta$ the contact angle.  The Lattice Boltzmann method can simulate 
accurately this phenomenon \cite{capillary}. Equation \ref{eq:z} predicts that the depth of the front inside 
the capillary varies linearly with time. A typical capillary system is illustrated in Fig.~\ref{fig:capillary}.
The behavior of the invading front in the presence of surface reaction is shown in Fig.~\ref{fig:front}.
There appears that the thickening of the surface behind the contact line retards the infiltration process, leading
also to pore closure. For narrower interstices, the retardation becomes important near the onset of pore closure,
so that this phenomenon is quite abrupt. For larger widths the resistance increases gradually. It is also interesting to
note that, for longer capillaries, the retardation effect is less marked but the infiltration process is slower 
(cf.~Ref.~\cite{capillary}). 

In order to shed more light on the process of pore closure, we consider the quantity $\Delta e$ defined
as the maximal width of the solid phase. Values close to $H/2$ indicate that the phenomenon of pore closure
occurred. Figure \ref{fig:solid} shows the time dependence of $\Delta e$ for different lengths of the capillaries 
and for different reaction-rate constants. It is important to remark that, for a given reaction-rate constant, the 
curves are quite similar. Indicatively, Eq.~\ref{eq:z} suggests that the infiltration velocity is reduced by a
factor of $2$ in passing from $L=500$ [lu] to $L=1'000$ [lu]. Altogether, it turns out that the kinetics of
surface growth is marginally influenced by the infiltration velocity. This means that narrow interstices are 
particularly detrimental for reactive infiltration because they are associated with smaller velocities and they 
occlude sooner and abruptly.

In Fig.~\ref{fig:radii} we compare the radii characterizing the infiltration process. $r_{\mathrm{min}}$ is the
minimum radius, $r_{\mathrm{h}}$ the hydraulic radius, $r_{\mathrm{ave}}$ the average radius and $r_{\mathrm{eff}}$ the
effective radius. The hydraulic radius is determined by means of the formula $r_{\mathrm{h}}=A/P$ \cite{dullien}; 
$A$ is the capillary area and $P$ the perimeter of the profile of the lower wall. The radius $r_{\mathrm{eff}}$
is obtained from Eq.~\ref{eq:z}: for $\cos\theta$ we use the results in the absence of reaction and the
infiltration velocity is estimated using the method of least squares. Of course, $r_{\mathrm{eff}}$ includes
the retardation effects due to surface growth and describes the whole infiltration process. For this reason, 
the other radii are rather larger at the beginning of the simulation and smaller towards the end. Furthermore,
we omit the results for the capillaries that occlude ($H=25$ [lu]) since the comparison is not particularly
speaking. In general, it can be seen that $r_{\mathrm{min}}$, $r_{\mathrm{h}}$ and $r_{\mathrm{ave}}$ vary linearly
with time. Deviations from the effective radius $r_{\mathrm{eff}}$ are smaller for the hydraulic radius, with
a few exceptions if $H=100$ [lu]. The hydraulic and average radii differ especially in the second half of
the simulations (see Fig.~\ref{fig:radii}). 


\section{Conclusions}

In this work, we could single out simple criteria allowing to ease the capillary infiltration 
of molten Si into C preforms. Namely, there appears that ideally the porous matrix should be
characterized by wide pathways as straight as possible. Importantly, the process of surface growth leading 
to pore closure is marginally influenced by the infiltration velocity. As a consequence, narrow
interstices are especially detrimental: their smaller size is in no way beneficial in order
to prevent the phenomenon of pore closure. Finally, our study suggests that the porosity characteristics 
of the carbon preforms are critical for the process of capillary infiltration. Future work will be devoted to this issue.

\acknowledgments

The research leading to these results has received funding from the European
Union Seventh Framework Programme (FP7/2007-2013) under grant agreement
n$^{\circ}$ 280464, project "High-frequency ELectro-Magnetic technologies
for advanced processing of ceramic matrix composites and graphite expansion''
(HELM).

\end{document}